\documentclass[conference]{IEEEtran}

\usepackage{amsmath,amssymb,amsthm}
\usepackage{cite}

\newcommand{\bA}{\mathbf{A}}

\newcommand{\bI}{\mathbf{I}}
\newcommand{\bP}{\mathbf{P}}
\newcommand{\bQ}{\mathbf{Q}}
\newcommand{\bU}{\mathbf{U}}
\newcommand{\bV}{\mathbf{V}}
\newcommand{\bX}{\mathbf{X}}
\newcommand{\bY}{\mathbf{Y}}

\newcommand{\bg}{\mathbf{g}}
\newcommand{\bh}{\mathbf{h}}
\newcommand{\bu}{\mathbf{u}}
\newcommand{\bv}{\mathbf{v}}
\newcommand{\bx}{\mathbf{x}}
\newcommand{\by}{\mathbf{y}}
\newcommand{\bz}{\mathbf{z}}
\newcommand{\bp}{\mathbf{p}}
\newcommand{\bq}{\mathbf{q}}
\newcommand{\bzero}{\mathbf{0}}

\newcommand{\FF}[1]{\mathbb{F}_{\hspace{-1pt}#1}}

\newtheorem{theorem}{Theorem}
\newtheorem{proposition}[theorem]{Proposition}
\newtheorem{corollary}[theorem]{Corollary}

\long\def\symbolfootnote[#1]#2{\begingroup
\def\thefootnote{\fnsymbol{footnote}}\footnote[#1]{#2}\endgroup}

\title{Repairing Multiple Failures in the Suh-Ramchandran Regenerating Codes}

\author{\IEEEauthorblockN{Junyu Chen}
\IEEEauthorblockA{Department of Information Engineering\\
the Chinese University of Hong Kong\\
Email: cj012@ie.cuhk.edu.hk}
\and
\IEEEauthorblockN{Kenneth W. Shum}
\IEEEauthorblockA{Institute of Network Coding\\
the Chinese University of Hong Kong\\
Email: wkshum@inc.cuhk.edu.hk}
}

\begin{document}

\maketitle

\begin{abstract} \symbolfootnote[0]{
This work was partially supported by a grant from the University Grants Committee of the Hong Kong Special Administrative Region, China (Project No. AoE/E-02/08).}
Using the idea of interference alignment, Suh and Ramchandran constructed a class of minimum-storage regenerating codes which can repair one systematic or one parity-check node with optimal repair bandwidth. With the same code structure, we show that in addition to single node failure, double node failures can be repaired collaboratively with optimal repair bandwidth as well. We give an example of how to repair double failures in the Suh-Ramchandran regenerating code with six nodes, and give the proof for the general case.
\end{abstract}

\smallskip


\section{Introduction}
In a distributed storage system, we encode and distribute a data file of size $B$ to $n$ storage nodes, with two properties that (i)
any $k$ nodes are sufficient in rebuilding the original file, and (ii) upon the failure of one or more storage nodes, we can recover the lost information efficiently. Property (i) is called the {\em $(n,k)$ recovery property}. We say that a coding scheme satisfies the {\em maximal-distance separable (MDS) property} if the $(n,k)$ recovery property is satisfied and each node stores $B/k$ units of data. The MDS property can be achieved by conventional MDS codes such as the Reed-Solomon (RS) codes. However, the communication and traffic required in repairing a failed node is very large if RS codes are employed, as the whole file must be downloaded before we re-encode the lost data in the failed node. The amount of traffic, measured in the number of packets transmitted from the surviving nodes to the new node, is coined {\em repair bandwidth} by Dimakis {\em et al.} in~\cite{DGWR07}. A lower bound on repair bandwidth is derived in the same work. A coding scheme with repair bandwidth attaining the lower bound is called a {\em regenerating code}.

The repair of failed storage nodes can be carried out in two ways. In the first one, called {\em exact repair}, the contents of the new nodes are exactly the same as the failed ones. The second is {\em functional repair}, in which the content need not be recovered exactly, but the $(n,k)$ recovery property is maintained. Exact repair has the advantage that we can store the data file in an uncoded form in some nodes, called the {\em systematic nodes}, while the other nodes store the parity-check data. In case we want to look up a small portion of the data file, we can connect to the node which holds that particular portion, without downloading the whole file. There are several existing constructions of regenerating codes for exact repair. One approach is to apply idea from {\em interference alignment} \cite{CadambeJafar, MaddahAli}, which is a concept in wireless communication for characterizing the degree of freedom of a wireless network. The regenerating code by Suh and Ramchandran~\cite{SK11} is one class of regenerating code constructed using this technique.

The Suh-Ramchandran code is designed for repairing single failure. For multiple failures, it was shown by Hu {\em et al.} in~\cite{HXWZL10} that by enabling data exchange, the repair bandwidth per new node can be further reduced. Suppose that we want to repair $r$ failures simultaneously. The repair process is divided into two phases.  In the first phase, each newcomer downloads $\beta_1$ packets from a set of $d$ surviving nodes. The system parameter $d$ is often called the {\em repair degree}. In the second phase, each pair of newcomers exchange $\beta_2$ packets in both directions. A regenerating code which repairs multiple-node failure jointly according to this two-phase protocol will be referred to as {\em cooperative} or {\em collaborative} regenerating code. The repair bandwidth per new node is denoted by $\gamma = d \beta_1 + (r-1) \beta_2$.

It was shown in~\cite{KSS} that for any cooperative regenerating code satisfying the MDS property, the repair bandwidth is lower bounded by
\begin{equation}
\gamma \geq \frac{B(d+r-1)}{ k(d+r-k) }.
\label{eq:MSCR}
\end{equation}
A cooperative regenerating code satisfying the MDS property and \eqref{eq:MSCR} with equality is called minimum-storage cooperative regenerating (MSCR) code. On the other hand, if the storage in each node is allowed to be larger than $B/k$, then the repair bandwidth of a cooperative regenerating code is lower bounded by
\begin{equation}
\gamma \geq \frac{B(2d+r-1)}{ k(2d+r-k) }.
\label{eq:MBCR}
\end{equation}
A cooperative regenerating code satisfying \eqref{eq:MBCR} with equality is called minimum-bandwidth cooperative regenerating (MBCR) code. When $r=1$, the bounds in \eqref{eq:MSCR} and \eqref{eq:MBCR} reduce to those for single-node repair in~\cite{DGWR07}.

There are some existing constructions of exact-repair MSCR and MBCR codes~\cite{Shum11, ShumHu11b, Jiekak, WangZhang, LeScouarnec12}. These constructions are summarized in Table~\ref{table}. For functional repair, the tradeoff curve between repair bandwidth and storage is derived in~\cite{ShumHu13}.

\begin{table}
\caption{Explicit constructions of cooperative regenerating codes.}
\begin{center}
\begin{tabular}{|c|c|c|} \hline
type & parameters & reference\\ \hline \hline
MSCR & $n \geq d+r$, $k=d$ & \cite{Shum11} \\ \hline
MBCR & $n=d+r$, $d=k$ & \cite{ShumHu11b} \\ \hline
MBCR & $n=d+r$, $d\geq k$, $r\geq 1$ & \cite{Jiekak} \\ \hline
MBCR & $n\geq d+r$, $d\geq k$, $r\geq 1$ & \cite{WangZhang} \\ \hline
MSCR & $n = d+r$,  $d\geq k$, $k=2$, $r=2$ & \cite{LeScouarnec12} \\ \hline
MSCR & $n =d+r = 2k$, $k\geq 3$, $r=2$& this paper \\ \hline
\end{tabular}
\end{center}
\label{table}
\end{table}

%One of the extreme case MBCR (minimum bandwidth cooperative regeneration) is contructed in \cite{ShumHu11b} for $d=k, n=d+r$ case and \cite{Jiekak} for $d \geq k, n=d+r$ case. Exact MBCR code for all possible values of $n,k,d,r$ is constructed in \cite{WangZhang}. Another extreme case MSCR (minimum storage cooperative regeneration) is contructed in \cite{Shum11} for $d=k,n=d+r$ case and \cite{LeScouarnec12} for $k=2, d \geq k$. The case of exact scalar MSCR code with $k \geq 3, d>k$ is shown to be impossible in \cite{LeScouarnec12}.

The objective of this paper is to show that the structure of the Suh-Ramchandran regenerating code also supports multiple-node repair. This disproves the assertion in \cite{LeScouarnec12} that ``it is not possible to repair exactly MSCR code with $k\geq 3$, $r\geq 2, d>k$ in the scalar case.'' After reviewing the Suh-Ramchandran construction in Section~\ref{sec:SR}, we state the main result of this paper in Section~\ref{sec:main}. In Section~\ref{sec:example}, an example with $(n, k) = (6, 3)$ is given.
The proof of the main theorem is stated in Section~\ref{sec:proof}.

\section{The Suh-Ramchandran Construction}
\label{sec:SR}
In the Suh-Ramchandran construction, the number of nodes, $n$, can be any integer larger than or equal to $2k$. For the ease of presentation, we focus on the case $n=2k$ in this paper.
We will use notations different from those in~\cite{SK11}, in order to emphasize the symmetry of the code, which will be crucial in the derivation of multiple-node recovery process.

Let $\FF{q}$ denote a finite field of size $q$. Each data symbol is regarded as a finite field element, and we will use a symbol as a unit of data. A symbol will also be called a {\em packet}.
The data file is divided into many data chunks, each containing $B=k^2$ symbols. All data chunks are encoded and treated in the same way. Hence,
we only need to describe the operations on one data chunk, and without loss of generality, we can assume that the data file consists of exactly $k^2$ symbols.

The construction requires four non-singular $k\times k$ matrices $\bU = [u_{ij}]$, $\bV= [v_{ij}]$, $\bP= [p_{ij}]$ and $\bQ = \bP^{-1} = [q_{ij}]$ over $\FF{q}$, satisfying
\begin{equation}
 \bU = \bV \bP \text{ and } \bV = \bU \bQ.
 \label{eq:UVPQ}
\end{equation}
Denote the columns of $\bU$ by $\bu_1, \bu_2,\ldots, \bu_k$, and the columns of $\bV$ by $\bv_1, \bv_2,\ldots, \bv_k$.
The columns of $\bU$ and $\bV$ are regarded as bases of $\FF{q}^k$, and the matrices $\bP$ and $\bQ$ are the change-of-basis matrices; the transformations in~\eqref{eq:UVPQ} are equivalent to
\begin{align*}
\bu_i &=  p_{1i} \bv_1 + p_{2i} \bv_2 + \cdots +  p_{ki} \bv_k , \\
\bv_i &=  q_{1i} \bu_1 + q_{2i} \bu_2 + \cdots + q_{ki} \bu_k ,
\end{align*}
for $i=1,2,\ldots, k$. Let
\begin{equation}
\hat\bU := (\bU^t)^{-1} \text{ and } \hat\bV := (\bV^t)^{-1},
\end{equation}
where the superscript $^t$ denotes the transpose operator. The columns of $\hat\bU$ (resp. $\hat\bV$) form the dual basis of $\bu_1, \bu_2,\ldots, \bu_k$ (resp. $\bv_1, \bv_2, \ldots, \bv_k$). Let the columns of $\hat\bU$ be $\hat\bu_1, \hat\bu_2,\ldots, \hat\bu_k$, and the columns of $\hat\bV$ be $\hat\bv_1, \hat\bv_2,\ldots, \hat\bv_k$.

Each node stores a column vector of length $k$ over $\FF{q}$. For $i=1,2,\ldots, k$, let the vector stored in node $i$ be denoted by $\bx_i$, and the vector stored in node $k+i$ be $\by_i$. Let $\bX$ (resp. $\bY$) be the $k\times k$ matrix whose columns are $\bx_i$ (resp. $\by_i$).

The Suh-Ramchandran regenerating code can be constructed in two ways.
In the first way, the data packets stored in nodes 1 to $k$ are uncoded symbols, and the packets stored in nodes $k+1$ to $2k$ are obtained by some linear transformation on the packets in nodes 1 to $k$, i.e., nodes 1 to $k$ are the systematic nodes, and nodes $k+1$ to $2k$ are the parity-check nodes. The parity-check symbols in nodes $k+1$ to $n$ are generated by
\begin{equation}
 \bY = \delta \hat\bV \bX^t \bU + \epsilon \bX \bP.
 \label{eq:primal}
\end{equation}
The variable $\delta$ and $\epsilon$ are elements in $\FF{q}$ to be determined later. If we let
$$
\bz_j  := \sum_{\ell=1}^k p_{\ell j} \bx_\ell
$$
to be the $j$-th column in matrix $\bX \bP$, then we write \eqref{eq:primal} in an alternate way as
\begin{equation}
 \by_j =  \Big(\delta \sum_{i=1}^k   \hat\bv_i \bu_j^t \bx_i \Big) + \epsilon  \bz_j,
 \tag{\ref{eq:primal}'}
\end{equation}
for $j=1,2,\ldots, k$.

In the second way of constructing the Suh-Ramchandran regenerating code, the packets in nodes $k+1$ to $2k$ are treated as information packets, while the packets in nodes 1 to $k$ are parity-check packets.
The matrix $\bX$ is obtained from $\bY$ by
\begin{equation}
 \bX = \delta' \hat\bU \bY^t \bV + \epsilon' \bY \bQ,
 \label{eq:dual}
\end{equation}
where  $\delta'$ and $\epsilon'$ are elements in $\FF{q}$.
We use the notation
$$
\bz_j' := \sum_{\ell=1}^k q_{\ell j} \by_\ell
$$
to denoted the $j$-th column of matrix $\bY \bQ$.
For $j=1,2, \ldots, k$,  the data stored in node $j$ can be expressed as
\begin{equation}
 \bx_j =  \Big( \delta' \sum_{i=1}^k \hat\bu_i \bv_j^t \by_i \Big) +  \epsilon' \bz_j'.
 \tag{\ref{eq:dual}'}
\end{equation}

The equivalence of these two ways of encoding is shown in the next theorem.

\begin{theorem}
Let $F(\bX) = \delta \hat\bV \bX^t \bU + \epsilon \bX \bP$ and $G(\bY) = \delta' \hat\bU \bY^t \bV + \epsilon' \bY \bQ$ be linear transformations from the vector space of $k\times k$ matrices to itself. If we choose $\delta$, $\delta'$, $\epsilon$ and $\epsilon'$ such that
\begin{align}
\delta \delta' + \epsilon \epsilon' &= 1, \text{ and } \label{eq:de1}\\
 \epsilon \delta' +  \delta\epsilon' &= 0, \label{eq:de2}
\end{align}
then the compositions $F\circ G$ and $G\circ F$ are the identity transformation.
\end{theorem}

\begin{proof} For all $k\times k$ matrices $\bX$, we have
\begin{align*}
G(F(\bX))
&= \delta' \hat\bU (\delta \bU^t \bX \hat\bV^t
+ \epsilon \bP^t \bX^t) \bV \\
& \qquad + \epsilon' (\delta \hat\bV \bX^t \bU + \epsilon \bX\bP) \bQ\\
&= (\delta \delta' + \epsilon \epsilon') \bX + ( \epsilon\delta' +  \delta\epsilon') \hat\bV \bX^t \bV = \bX.
\end{align*}
The proof of $F(G(\bY))=\bY$ is similar.
\end{proof}

In \cite{SK11}, Suh and Ramchandran prove the following.

\begin{theorem}[\cite{SK11}]
The Suh-Ramchandran regenerating codes satisfies the MDS property if all square submatrices of matrix $\bP$ are non-singular.
\label{thm:MDS}
\end{theorem}

We will call a matrix {\em super-regular} if all square submatrices are non-singular. It can be proved that the inverse of a super-regular matrix is also super-regular. Therefore in Theorem~\ref{thm:MDS}, it is equivalent to pick the matrix $\bQ$ to be super-regular.
%A well-known class of super-regular matrices are Cauchy matrices, which are square matrices of the form
%$[ (a_i - b_j)^{-1} ]$
%with distinct $a_i$'s and $b_j$'s.

\section{Main Result}
\label{sec:main}
The main result of this paper is to show that the Suh-Ramchandran regenerating code, which is originally aiming at repairing single-node failure, can repair the following patterns of multiple-node failures with minimal repair bandwidth.

\begin{theorem}
Suppose that in the Suh-Ramchandran construction, the parameters $\bV$, $\bP$, $\epsilon$, $\delta$, $\epsilon'$ and $\delta'$ are chosen such that
\begin{itemize}
\item $\bV$ is a $k\times k$ non-singular matrices over $\FF{q}$,
\item $\bP$ is a $k\times k$ super-regular matrices over $\FF{q}$,
\item $\epsilon$, $\delta$, $\epsilon'$ and $\delta'$ are non-zero and  satisfy \eqref{eq:de1} and \eqref{eq:de2},
\item $p_{ij} q_{ji} \neq 1$ for all $i$ and $j$ in $\{1,2,\ldots, k\}$.
\end{itemize}
%where $q_{ji}$ is the $(j,i)$-entry of $\bP^{-1}$.
Then we can exactly repair
\begin{itemize}
\item $r$ systematic nodes,  for any $r$ between 1 and $k$,
\item $r$ parity-check nodes, for any $r$ between 1 and $k$,
\item any pair of systematic node and parity-check node,
\end{itemize}
with repair bandwidth attaining the lower bound in~\eqref{eq:MSCR} and repair degree $d$ equal to $n$ minus the number of failed nodes repaired cooperatively.
\label{thm:main}
\end{theorem}

The proof of Theorem~\ref{thm:main} is given in Section~\ref{sec:proof}.

We need to choose the coding coefficients such that the conditions in Theorem~\ref{thm:main} are satisfied. First of all,
if we square both sides of \eqref{eq:de1} and \eqref{eq:de2} and subtract, we get
$$(\delta^2 - \epsilon^2)((\delta')^2 - (\epsilon')^2) = 1.$$
Hence, we have $\delta^2 \neq \epsilon^2$ and $(\delta')^2 \neq (\epsilon')^2$.
As the determinant of the $2\times 2$ matrix in
\begin{equation}
 \begin{bmatrix}
 \delta & \epsilon \\ \epsilon & \delta
  \end{bmatrix}
  \begin{bmatrix} \delta' \\ \epsilon' \end{bmatrix}
  = \begin{bmatrix} 1 \\ 0 \end{bmatrix}
  \label{eq:de_matrix}
\end{equation}
is necessarily non-zero, we can choose $\delta$ and $\epsilon$ to be a pair of nonzero elements in $\FF{q}$ such that $\delta^2\neq \epsilon^2$, and then obtain $\epsilon'$ and $\delta'$
by solving~\eqref{eq:de_matrix}. The values of $\epsilon'$ and $\delta'$ so obtained are provably non-zero.

%For example, we can take $\delta=\delta'=\sqrt{\theta}$, $\epsilon=\sqrt{\theta-1}=-\epsilon'$, for some element $\theta\in\FF{q}$ provided that the square roots of $\theta$ and $\theta-1$ are well-defined in~$\FF{q}$.

Secondly,
for a Cauchy matrix $\bP = [(a_i - b_j)^{-1}]$,
the  $(j,i)$-entry of $\bP^{-1}$ can be calculated by
\begin{equation}
 q_{ji} = (a_i - b_j)\frac{\prod_{\ell\neq i} (b_j-a_\ell)}{\prod_{\ell\neq i} (a_i-a_\ell)} \cdot \frac{\prod_{\ell\neq j} (a_i-b_\ell)}{\prod_{\ell\neq j} (b_j-b_\ell)}.
 \label{eq:cofactor}
\end{equation}
See for example~\cite{Schechter} for a derivation of~\eqref{eq:cofactor}. Whence, the condition $p_{ij} q_{ji} \neq 1$ is equivalent to
\[
\prod_{\ell\neq i} (b_j-a_\ell) \cdot \prod_{\ell\neq j} (a_i-b_\ell) -
\prod_{\ell\neq i} (a_i-a_\ell) \cdot \prod_{\ell\neq j} (b_j-b_\ell) \neq 0.
\]
Let $F_{ij}$ be the left-hand side of the above equation, regarded as a mutli-variate polynomial in $a_i$'s and $b_j$'s.
Constructing a Cauchy matrix $\bP$ satisfying the conditions in Theorem~\ref{thm:main} amounts to finding $a_i$'s and $b_j$'s such that the product $\prod_{1\leq i,j \leq k} F_{ij}$
is evaluated to a non-zero constant in~$\FF{q}$. By Schwartz-Zippel lemma (see e.g.~\cite[Corollary 19.18]{Raymond08}), this can be done if the  finite field size $q$ is sufficiently large.

\begin{corollary}
With sufficiently large finite field $\FF{q}$, we can repair single and double node failures in the Suh-Ramchandran regenerating code with optimal repair bandwidth.
\end{corollary}

%This disprove the assertion in~\cite{LeScouarnec12} that ``it is not possible to repair exactly MSCR code with $k\geq 3$ and $r\geq 2$ in the scalar case, such that each node stores $\alpha = d-k+r$ packets.''

\section{An Example for $n=6$ and $k=3$}
\label{sec:example}
In this section, we illustrate how to repair two node failures in the rate-$1/2$ Suh-Ramchandran code for $n=6$ nodes.

\smallskip
{\bf Encoding.}  There are $B=9$ symbols to be encoded and distributed to $n=6$ storage nodes. Let us agree that the first three nodes are systematic nodes, and the last three nodes are parity-check nodes. Each node stores a column vector of length~3. We let $\bV = [ \bv_1 | \bv_2 | \bv_3 ]$ be a non-singular $3\times 3$ matrices, and
$\bP = [p_{ij}]_{i,j=1}^3$
%\begin{bmatrix}
%p_{11} & p_{12} & p_{13} \\
%p_{21} & p_{22} & p_{23} \\
%_{31} & p_{32} & p_{33}
%\end{bmatrix}
be a Cauchy matrix, so that the MDS property is guaranteed by Theorem~\ref{thm:MDS}. Let $\bU = [ \bu_1 | \bu_2 | \bu_3 ] = \bV \bP$ and  denote the inverse of $\bP$ by
$\bQ =
\bP^{-1} = [q_{ij}]_{i,j=1}^3$.
%\begin{bmatrix}
%q_{11} & q_{12} & q_{13} \\
%q_{21} & q_{22} & q_{23} \\
%q_{31} & q_{32} & q_{33}
%\end{bmatrix}.

The encoding is illustrated in the following table:
\[
\begin{array}{|c|c|} \hline
\text{Node } & \text{Content} \\ \hline \hline
1 & \bx_1 \\ \hline
2 & \bx_2 \\ \hline
3 & \bx_3 \\ \hline
4 &  \phantom{\Big(}\by_1=\delta \sum_{j=1}^3\hat\bv_j \bu_1^t \bx_j  +\epsilon \bz_1
\\ \hline
5 & \phantom{\Big(}\by_2=\delta \sum_{j=1}^3\hat\bv_j \bu_2^t \bx_j  + \epsilon \bz_2
\\ \hline
6 & \phantom{\Big(} \by_3=\delta \sum_{j=1}^3\hat\bv_j \bu_3^t \bx_j  + \epsilon \bz_3 \\ \hline
\end{array}
\]

{\bf Repair.} Upon the failure of two storage nodes, each surviving node sends a linear combination of the stored symbols to each of the failed node. The first phase of the repair procedure is as follows.
\begin{enumerate}
\item
If node $i$ is one of the failed node, for $i=1,2,3$, a surviving node takes the inner product of the stored vector and $\bv_i$, and sends it to newcomer $i$.
\item If node $3+j$ is one of the failed node, for $j=1,2,3$, a surviving node takes the inner product of the stored vector and $\bu_j$, and sends it to newcomer~$3+j$.
\end{enumerate}

By the symmetry of the code structure, it is sufficient to discuss the repair of (i) two parity-check nodes, and (ii) one systematic node and one parity check node.

\smallskip
{\bf Repair of two parity-check nodes.}
Without loss of generality, we consider the repair of nodes 4 and~5.
After the first phase of the repair process, newcomer 4 receives four symbols,
$$\bu_1^t \bx_1,\
\bu_1^t \bx_2,\
\bu_1^t \bx_3 \text{ and }
\bu_1^t \by_3 = \delta \bu_3^t \bz_1 +
\epsilon	\bu_1^t \bz_3.$$
The symbols received by newcomer 5 are
$$\bu_2^t \bx_1,\
\bu_2^t \bx_2, \
\bu_2^t \bx_3 \text{ and }
\bu_2^t \by_3 =\delta \bu_3^t \bz_2+
\epsilon	\bu_2^t \bz_3.$$

Recall that newcomer 5 wants to compute
\begin{equation}
\by_2 =
\delta(\hat\bv_1 \bu_2^t \bx_1 + \hat\bv_2 \bu_2^t \bx_2 + \hat\bv_3 \bu_2^t \bx_3) +  \epsilon \bz_2.
\label{eq:y2}
\end{equation}
The first term can be obtained from $\bu_2^t \bx_1$, $\bu_2^t \bx_2$ and $\bu_2^t \bx_3$. For the second term, newcomer 5 first calculates
\begin{align*}
 \bu_2^t \bz_2 &= p_{12} \bu_2^t \bx_1 + p_{22} \bu_2^t \bx_2 + p_{32} \bu_2^t \bx_3 , \\
 \bu_3^t \bz_2 &= \frac{1}{\delta} \Big( \bu_2^t \by_3 -
 \epsilon p_{13} \bu_2^t \bx_1 - \epsilon p_{23} \bu_2^t \bx_2 - \epsilon p_{33} \bu_2^t \bx_3  \Big).
\end{align*}
and then asks newcomer 4 for a copy
$$\bu_1^t \bz_2 = p_{11} \bu_1^t \bx_1 + p_{21} \bu_1^t \bx_2 + p_{31} \bu_1^t \bx_3,$$
which can be computed by newcomer 4.
In the computation of $\bu_3^t \bz_2$, it is obvious that we need to impose the condition that $\delta\neq 0$. Then,
by the linear independence of $\bu_1$, $\bu_2$ and $\bu_3$, newcomer~5 can  regenerating the second term in~\eqref{eq:y2}.

Similarly, newcomer 4 can regenerate $\by_1$
after newcomer 5 has sent $ \bu_2^t \bz_1$ to newcomer~4.

\smallskip
{\bf Repair of a systematic node and a parity-check node.}
Without loss of generality, we consider the repair of nodes 1 and 5.
After the first phase of the repair process, newcomer 1 receives $\bv_1^t \bx_2$, $\bv_1^t \bx_3$,
\begin{align*} \bv_1^t \by_1 &= \delta \bu_1^t \bx_1 + \epsilon \bv_1^t \bz_1, \text{ and }
\bv_1^t \by_3  = \delta \bu_3^t \bx_1 + \epsilon \bv_1^t \bz_3,
\end{align*}
while newcomer 5 receives $\bu_2^t \bx_2$,  $\bu_2^t \bx_3$,
\begin{align*}
\bu_2^t \by_1 &= \delta \bu_1^t \bz_2  + \epsilon \bu_2^t \bz_1, \text{ and }
\bu_2^t \by_3 = \delta \bu_3^t \bz_2  + \epsilon \bu_2^t \bz_3.
\end{align*}

Newcomer 5 computes a linear combination of the received symbols,
\begin{align*}
q_{11} \bu_2^t \by_1 + q_{31} \bu_2^t \by_3
  +  (\delta+\epsilon) [p_{22} q_{21} \bu_2^t \bx_2
  +p_{32}q_{21}  \bu_2^t \bx_3 ].
\end{align*}
The coefficients are chosen so that it can be simplified to
\begin{equation}
\delta \bv_1^t \bz_2 + (\epsilon - (\epsilon+\delta) p_{12} q_{21}) \bu_2^t \bx_1, \label{eq:4to1a}
\end{equation}
which is a linear combination of $\bv_1^t \bz_2$ and $\bu_2^t \bx_1$.
(We have used the orthogonality relation $\sum_{\ell} p_{i\ell} q_{\ell j}$ is equal to the Kronecker delta function $\delta_{ij}$.)
In the second phase of the repair process, newcomer~5 sends the symbol in~\eqref{eq:4to1a} to newcomer~1.

Since newcomer 1 knows $\bv_1^t \bx_2$ and $\bv_1^t \bx_3$, newcomer 1 can compute
$$\big( \delta p_{12} \bv_1^t + (\epsilon - (\epsilon+\delta) p_{12} q_{21} ) \bu_2^t \big) \bx_1
$$
by subtracting $\delta p_{22} \bv_1^t \bx_2$ and $\delta p_{32} \bv_1^t \bx_3$. Next, newcomer 1 calculates
\begin{align*}
\bv_1^t \by_1 - \epsilon p_{21} \bv_1^t \bx_2 - \epsilon p_{31} \bv_1^t \bx_3 &=(\delta \bu_1^t + \epsilon p_{11}  \bv_1^t) \bx_1, \text{ and}\\
\bv_1^t \by_3 - \epsilon p_{23} \bv_1^t \bx_2 - \epsilon p_{33} \bv_1^t \bx_3 &=(\delta \bu_3^t + \epsilon p_{13}  \bv_1^t) \bx_1.
\end{align*}
The vector $\bx_1$ can be recovered if the matrix
\[
\begin{bmatrix}
(\epsilon - (\epsilon + \delta) p_{12} q_{21} )\bu_2^t +  \delta p_{12} \bv_1^t \\
\delta \bu_1^t + \epsilon p_{11}  \bv_1^t \\
\delta \bu_3^t + \epsilon p_{13}  \bv_1^t
\end{bmatrix}
\]
is non-singular.

Using the symmetry of the code, newcomer 5 can recover the lost information in a similar way.

\section{Proof of the Main Theorem}
\label{sec:proof}
We use the first encoding method of the Suh-Ramchandran code;
the entries in $\bX$ are the source symbols and the entries in $\bY$ are the parity-check symbols calculated by \eqref{eq:primal}.
In the first phase of the repair procedure, the packet sent from a surviving node to a newcomer is computed as follows:
\begin{enumerate}
\item
If node $i$ is one of the failed node, for $i=1,2,\ldots, k$, then a surviving node takes the inner product of the stored vector and $\bv_i$, and sends it to newcomer $i$.
\item If node $k+i$ is one of the failed node, for $i=1,2,\ldots, k$, then a surviving node takes the inner product of the stored vector and $\bu_i$, and sends it to newcomer~$k+i$.
\end{enumerate}

\smallskip
{\bf Repair of $r$ parity-check or systematic nodes, $1 \leq r \leq k$.}

By the symmetry between $\bX$ and $\bY$, it suffices to consider the repair of parity-check nodes.
%Let
%$$
%\bz_j = \sum_{l=1}^k p_{lj} \bx_l
%$$

Suppose that nodes $k+1$ to $k+r$ fail. For $i=1,2,\ldots, r$,
the symbols received by newcomer $k+i$ are $\bu_i^t \bx_1$ to $\bu_i^t \bx_k$ and $\bu_i^t \by_j = \delta \bu_j^t \bz_i +
\epsilon	\bu_i^t  \bz_j$ for  $j=r+1,r+2,\ldots,k$.

Recall that newcomer $k+i$ wants to regenerate
\[
\Big( \delta \sum_{\ell=1}^k \hat\bv_\ell \bu_i^t \bx_\ell \Big) +\epsilon \bz_i.
\]
The first term is known to newcomer $k+i$ after the first phase, and
can be reconstructed from $\bu_i^t \bx_1$ to  $\bu_i^t \bx_k$. For the second term, newcomer $k+i$ calculates
\begin{align*}
\bu_i^t \bz_i & = \sum_{\ell=1}^k p_{\ell i} \bu_i^t \bx_\ell, \  \ \text{ and}\\
\bu_j^t \bz_i &= \frac{1}{\delta}\bu_i^t \by_j - \frac{\epsilon}{\delta} \Big( \sum_{\ell=1}^k p_{\ell j} \bu_i^t \bx_\ell\Big),
\end{align*}
for $j=r+1, r+2,\ldots, k$, and asks the other $r-1$ newcomers each for a copy of
$\bu_j^t \bz_i$, for $1 \leq j \leq r, j \neq i$.
Using the fact that $\bu_1, \bu_2, \ldots, \bu_k$ are linearly independent, newcomer $k+i$ can then solve for $\bz_i$.

\smallskip
{\bf Repair of a systematic node and a parity-check node.}

Suppose nodes $a$ and $k+b$ fail, where $a$ and $b$ are integers between 1 and $k$. We want to replace them by newcomer $a$ and newcomer~$k+b$. Let $[k]$ denote $\{1,2,\ldots, k\}$.

After the first phase of the repair process, newcomer $a$ receives
\begin{align*}
\bv_a^t \bx_i & \text{ for } i\in [k]\setminus\{a\}, \text{ and } \\
\bv_a^t \by_j &=\delta \bu_j^t \bx_a + \epsilon \bv_a^t \bz_j  \text{ for } j\in[k]\setminus\{b\},
\end{align*}
and newcomer $k+b$ receives
\begin{align*}
\bu_b^t \bx_i & \text{ for } i\in [k]\setminus\{a\} , \text{ and} \\
\bu_b^t \by_j &= \delta \bu_j^t \bz_b + \epsilon \bu_b^t  \bz_j \text{ for } j \in [k]\setminus\{b\}.
\end{align*}
In the second phase, newcomer $k+b$ sends the linear combination
\begin{align*}
& \phantom{=} \sum_{j\neq b} q_{ja} \bu_b^t\by_j + (\delta+\epsilon)\sum_{i \neq a} p_{ib} q_{ba} \bu_b^t \bx_i \\
&= \delta \bv_a^t \bz_b + (\epsilon -(\epsilon+\delta)p_{ab}q_{ba}) \bu_b^t \bx_a,
\end{align*}
where $j$ runs over $[k]\setminus\{b\}$ and $i$ runs over $[k]\setminus \{a\}$, to newcomer $a$. Newcomer $a$ then calculates
$$\big( \delta p_{ab} \bv_a^t  + ( \epsilon-(\epsilon+\delta)p_{ab}q_{ba} \bu_b^t \big) \bx_a,
$$
and
$(\delta \bu_j^t + \epsilon p_{aj}  \bv_a^t) \bx_a$
for $j\in[k]\setminus\{b\}$.
The vector $\bx_a$ can be recovered if the following $k\times k$ matrix
\begin{equation}
\begin{bmatrix}
(\epsilon - (\epsilon + \delta) p_{ab} q_{ba} )\bu_b^t +  \delta p_{ab} \bv_a^t \\
\delta \bu_1^t + \epsilon p_{a1}  \bv_a^t \\
\vdots \\
\delta \bu_{b-1}^t + \epsilon p_{a, b-1}  \bv_a^t \\
\delta \bu_{b+1}^t + \epsilon p_{a, b+1}  \bv_a^t \\
\vdots \\
\delta \bu_k^t + \epsilon p_{ak}  \bv_a^t
\end{bmatrix}
\label{eq:14a}
\end{equation}
is non-singular. We will show in Prop.~\ref{prop:miracle} that the determinant of this matrix is non-zero if  $p_{ab} q_{ba} \neq 1$.

Using the symmetric of the code, newcomer $k+b$ can recover the lost information after receiving
\[
  \delta' \bu_b^t \bz_a' + (\epsilon'-(\epsilon'+\delta')q_{ba} p_{ab}) \bv_a^t \by_b
\]
from newcomer $a$, provided that $p_{ab} q_{ba} \neq 1$.

\begin{proposition}
Suppose that $\bV$, $\bP$,  $\epsilon$, $\delta$, $\epsilon'$ and $\delta'$ satisfy the criteria in Theorem~\ref{thm:main}.
Then the determinant of the matrix in \eqref{eq:14a} is non-zero.
\label{prop:miracle}
\end{proposition}

\begin{proof}
We divide the proof into two cases.

{\em Case 1:} $\epsilon - (\epsilon+\delta)p_{ab} q_{ba}=0$. In this case, we can row-reduce the matrix in \eqref{eq:14a} to
\[
 \begin{bmatrix}
 \delta p_{ab} \bv_a^t \\
 \delta \bU^{t}_{[k]\setminus\{b\}}
 \end{bmatrix}
 =
  \begin{bmatrix}
 \delta p_{ab} \sum_{\ell=1}^{k} q_{\ell a} \bu_\ell^t \\
 \delta \bU^{t}_{[k]\setminus\{b\}}
 \end{bmatrix}
\]
where $\bU_{[k]\setminus\{b\}}$ denotes the $k\times (k-1)$ matrix
$$\bU_{[k]\setminus\{b\}} =
 \begin{bmatrix}
 \bu_1 &
 \dots &
 \bu_{b-1} &
 \bu_{b+1} &
 \dots &
 \bu_k
 \end{bmatrix}.
$$
It can further be row-reduced to a non-singular matrix, and thus has non-zero determinant.

{\em Case 2:} $\epsilon - (\epsilon+\delta)p_{ab} q_{ba} \neq 0$.
After substituting $\bv_a$ by $\sum_{\ell=1}^{k} q_{\ell a} \bu_\ell$,  the matrix in \eqref{eq:14a} can be factored as
\begin{equation}
\begin{bmatrix}
\epsilon - \epsilon p_{ab} q_{ba} & \delta p_{ab} \bq_{[k]\setminus\{b\},a}^t \\
\epsilon q_{ba} \bp_{a,[k]\setminus\{b\}} & \delta \bI + \epsilon \bp_{a,[k]\setminus\{b\}} \bq_{[k]\setminus\{b\},a}^t \\
\end{bmatrix}
\begin{bmatrix}
\bu_b^t \\ \bU^{t}_{[k]\setminus\{b\}}
\end{bmatrix}
\label{eq:14e}
\end{equation}
where $\mathbf{I}$ is the $(k-1)\times(k-1)$ identity matrix, $\bp_{a,[k]\setminus\{b\}}$ is the column vector
$$\bp_{a,[k]\setminus\{b\}} :=[p_{a1}\ \cdots p_{a, b-1} \ p_{a, b+1} \cdots p_{ak}]^t,
$$
and
$\bq_{[k]\setminus\{b\},a}$ is the column vector
$$\bq_{[k]\setminus\{b\},a}:=[q_{1a}\ \cdots q_{b-1,a} \ q_{b+1,a} \cdots q_{ka}]^t.$$

The non-singularity of \eqref{eq:14a} is equivalent to the non-singularity of the first factor in \eqref{eq:14e}, which in turn can be decomposed as $$\bA+\bg \bh^{t},$$
where  
\begin{gather*}
  \bA =
  \begin{bmatrix}
  \epsilon - (\epsilon+\delta)p_{ab} q_{ba} & \bzero \\
  \bzero & \delta \bI \\
  \end{bmatrix} \\
\intertext{is a diagonal matrix, and}  
  \bg =
  \begin{bmatrix}
  \delta p_{ab} \\ \epsilon \bp_{a,[k]\setminus\{b\}}
  \end{bmatrix}, \
  \bh =
  \begin{bmatrix}
   q_{ba} \\ \bq_{[k]\setminus\{b\},a}
  \end{bmatrix}
\end{gather*}
are column vectors.

The first summand is non-singular because $\epsilon - (\epsilon+\delta)p_{ab} q_{ba}$ and $\delta$ are non-zero. By the Sherman-Morrison formula~\cite[p.18]{HornJohnson}, we see that the matrix in \eqref{eq:14a} is invertible if
$$1 + \bh^{t} \bA^{-1} \bg$$
is non-zero. Using the identity $\sum_{\ell=1}^{k} p_{a\ell}q_{\ell a}=1$, the above expression can be simplified to
\[
\frac{\epsilon (\epsilon+\delta)(1-p_{ab}q_{ba})^2 }{ \epsilon - (\epsilon+\delta)p_{ab} q_{ba}},
\]
which is nonzero because $\epsilon \neq 0$, $\delta^2 \neq \epsilon^2$, and $p_{ab} q_{ba} \neq 1$.
\end{proof}

\section{Concluding Remarks}

In this paper we show that with the regenerating code constructed by Suh and Ramchandran, which is originally designed for repairing any single node failure, multiple-node failures can also be repaired cooperatively with optimal repair bandwidth. Indeed, we can repair any set of systematic nodes, any set of parity-check nodes, or any pair of nodes. However, the technique that we used in this paper cannot be extended to the optimal repair of one systematic node and two parity-check nodes.

%\bibliographystyle{IEEEtran}

%\bibliography{DStorage}

\begin{thebibliography}{10}
\providecommand{\url}[1]{#1}
\csname url@samestyle\endcsname
\providecommand{\newblock}{\relax}
\providecommand{\bibinfo}[2]{#2}
\providecommand{\BIBentrySTDinterwordspacing}{\spaceskip=0pt\relax}
\providecommand{\BIBentryALTinterwordstretchfactor}{4}
\providecommand{\BIBentryALTinterwordspacing}{\spaceskip=\fontdimen2\font plus
\BIBentryALTinterwordstretchfactor\fontdimen3\font minus
  \fontdimen4\font\relax}
\providecommand{\BIBforeignlanguage}[2]{{%
\expandafter\ifx\csname l@#1\endcsname\relax
\typeout{** WARNING: IEEEtran.bst: No hyphenation pattern has been}%
\typeout{** loaded for the language `#1'. Using the pattern for}%
\typeout{** the default language instead.}%
\else
\language=\csname l@#1\endcsname
\fi
#2}}
\providecommand{\BIBdecl}{\relax}
\BIBdecl

\bibitem{DGWR07}
A.~G. Dimakis, P.~B. Godfrey, M.~J. Wainwright, and K.~Ramchandran, ``Network
  coding for distributed storage system,'' in \emph{Proc. {IEEE} Int. Conf. on
  Computer Comm. (INFOCOM)}, Anchorage, Alaska, May 2007, pp. 2000--2008.

\bibitem{CadambeJafar}
V.~R. Cadambe and C.~Jafar, ``Interference alignment and degrees of freedom of
  the {K}-user interference channel,'' \emph{{IEEE} Trans. Inf. Theory},
  vol.~54, no.~8, pp. 3425--3441, Aug. 2008.

\bibitem{MaddahAli}
M.~A. Maddah-Ali, A.~S. Motahari, and A.~K. Khandani, ``Communication over
  {MIMO} {X} channels: Interference alignment, decomposition, and performance
  analysis,'' \emph{{IEEE} Trans. Inf. Theory}, vol.~54, no.~8, pp. 3457--3470,
  Aug. 2008.

\bibitem{SK11}
C.~Suh and K.~Ramchandran, ``Exact-repair {MDS} code construction using
  interference alignment,'' \emph{{IEEE} Trans. Inf. Theory}, vol.~57, no.~3,
  pp. 1425--1442, Mar. 2011.

\bibitem{HXWZL10}
Y.~Hu, Y.~Xu, X.~Wang, C.~Zhan, and P.~Li, ``Cooperative recovery of
  distributed storage systems from multiple losses with network coding,''
  \emph{{IEEE} J. on Selected Areas in Commun.}, vol.~28, no.~2, pp. 268--276,
  Feb. 2010.

\bibitem{KSS}
A.-M. Kermarrec, N.~{Le Scouarnec}, and G.~Straub, ``Repairing multiple
  failures with coordinated and adaptive regenerating codes,'' in \emph{Proc.
  Int. Symp. on Network Coding (Netcod)}, Beijing, Jul. 2011, pp. 88--93.

\bibitem{Shum11}
K.~W. Shum, ``Cooperative regenerating codes for distributed storage systems,''
  in \emph{IEEE Int. Conf. Comm. (ICC)}, Kyoto, Jun. 2011, pp. 1--5.

\bibitem{ShumHu11b}
K.~W. Shum and Y.~Hu, ``Exact minimum-repair-bandwidth cooperative regenerating
  codes for distributed storage systems,'' in \emph{Proc. {IEEE} Int. Symp.
  Inf. Theory}, St. Petersburg, Aug. 2011, pp. 1374--1378.

\bibitem{Jiekak}
S.~Jiekak and N.~{Le Scouarnec}, ``{CROSS-MBCR}: Exact minimum bandwidth
  coordinated regenerating codes,'' arXiv:1207.0854v1 [cs.IT], Jul. 2012.

\bibitem{WangZhang}
A.~Wang and Z.~Zhang, ``Exact cooperative regenerating codes with
  minimum-repair-bandwidth for distributed storage,'' arXiv:1207.0879v1
  [cs.IT], Jul. 2012.

\bibitem{LeScouarnec12}
N.~{Le Scouarnec}, ``Exact scalar minimum storage coordinated regenerating
  codes,'' in \emph{Proc. {IEEE} Int. Symp. Inf. Theory}, Cambridge, Jul. 2012,
  pp. 1197--1201.

\bibitem{ShumHu13}
K.~W. Shum and Y.~Hu, ``Cooperative regenerating codes,'' 2013, to appear in
  IEEE Trans. Inf. Theory.

\bibitem{Schechter}
S.~Schechter, ``On the inversion of certain matrices,'' \emph{Mathematical
  Tables and other Aids to Computation}, vol.~13, no.~66, pp. 73--77, 1956.

\bibitem{Raymond08}
R.~W. Yeung, \emph{Information theory and network coding}.\hskip 1em plus 0.5em
  minus 0.4em\relax New York: Springer, 2008.

\bibitem{HornJohnson}
R.~A. Horn and C.~R. Johnson, \emph{Matrix analysis}.\hskip 1em plus 0.5em
  minus 0.4em\relax Cambrdige: Cambridge University Press, 1985.

\end{thebibliography}


\end{document}